\documentclass{appolb}
\usepackage{amsfonts}
\usepackage{epsfig}

\begin{document}
\title{QUANTUM MECHANICS WITH NEUTRAL KAONS %
\thanks{Presented at the Final Euridice Meeting, Effective Theories of Colour
and Flavour: from EURODAPHNE to EURIDICE, Kazimierz (Poland), August 24--27, 2006.
Work also supported by INFN, MEC FIS2005-1369 and Consol\'{\i}der--Ingenio 2010, QOIT.}%
}
\author{A.~Bramon
\address{Grup de F{\'\i}sica Te\`orica,
Universitat Aut\`onoma de Barcelona,\\
E--08193 Bellaterra, Spain}
\and
G. Garbarino
\address{Dipartimento di Fisica Teorica, Universit\`a di Torino and
INFN, \\ Sezione di Torino, I--10125 Torino, Italy}
\and
B.~C.~Hiesmayr
\address{Institute for Theoretical Physics, Faculty of Physics, University of Vienna, \\
A--1090 Vienna, Austria}
}
\maketitle
\begin{abstract}
We briefly illustrate a few tests of quantum mechanics which can be
performed with entangled neutral kaon pairs at a $\Phi$--factory.
This includes a quantitative formulation of Bohr's complementarity
principle, the quantum eraser phenomenon and various forms of Bell
inequalities.
\end{abstract}

\section{Introduction}
Some of the {\it Gedanken}--experiments discussed in the early days
of Quantum Mechanics (QM) by its founding fathers have been recently
reanalyzed in their original form or in slightly modified versions.
These reanalyses have allowed an experimental confirmation of
the QM predictions and deep insights into the heart of QM. Good
examples are recent researches on  basic subjects such as
Bohr's complementarity principle or other subtle QM issues which
featured when the theory had already been completed, e.g., the so
called Einstein, Podolsky and Rosen paradox and the closely related
subject of Bell inequalities. Most of this progress has been
achieved thanks to the many advances in quantum optics and photonic
experiments, however, the improved technologies also allowed to
perform such tests with atoms and ions.

The purpose of the present contribution is to show that 
high energy physics systems, such as kaons, can also be considered 
to discuss basic questions
of QM and that sometimes they are more instructive and simple than
their photonic or atomic analogues.

We review a series of recent papers dealing with QM issues using
neutral kaons as the relevant quantum states. 
We start in Sect.~\ref{tnks} with a short summary of the required neutral kaon
formalism paying attention to the quantum measurements and to
entangled kaon states. In Sect.~\ref{quco} we discuss a quantitative
version of Bohr's complementarity for single neutral kaon states.
Entangled kaon pairs are then used to introduce a kaonic `quantum
eraser' in Sect.~\ref{qe} which may be tested at the
$\Phi$--factory. We then end with the subtle topic of Bell
inequalities with neutral kaons in Sect.~\ref{bit}.

\section{The neutral kaon system}
\label{tnks}
\subsection{Two bases: $\{K^0,\bar{K^0}\}$ and $\{K_S,K_L\}$}

Neutral kaons are pseudoscalar mesons consisting of a quark--antiquark bound state,
${K^0} \sim d\bar s$ and $\bar{K^0} \sim s \bar d$.
These two states define the `strangeness' or `strong--interaction' basis:
$|{K^0}\rangle$ and $|\bar{K^0}\rangle$ with strangeness
$S=+1$ and $S = -1$, respectively.  This is the suitable basis to
analyze $S$--conserving electromagnetic and strong interaction processes, such as the
creation of ${K^0}\bar{K^0}$ systems from non--strange initial states
(e.g., $e^+ e^- \to \phi (1020) \to {K^0} \bar{K^0} $ and
$p \bar p \to {K^0} \bar{K^0} $),
and the detection of neutral kaons via strong kaon--nucleon interactions.
This `strangeness' basis is orthonormal, $\langle K^0 | \bar{K^0}\rangle =0$.

Weak interaction phenomena allow for strangeness non--conservation thus
introducing new effects ---such as ${K^0}$--$\bar{K^0}$ oscillations---
as well as neutral kaon time evolution and decay.
All these phenomena, together with  kaon propagation in a medium with its
associated regeneration effects, require the use of other relevant bases.

The `free--space' basis, $\{K_S,K_L\}$, consists of the
so called $K$--short and $K$--long states which are the normalized eigenvectors
of the effective weak Hamiltonian $H_{\rm free}$ governing neutral kaon time evolution
in free--space:
\begin{equation}
\label{evol} i\frac{d}{d\tau}|K_{S,L}(\tau)\rangle=H_{\rm
free}\,|K_{S,L}(\tau)\rangle\, ,\,\,\, H_{\rm free} = \pmatrix {
\lambda_+ & \lambda_-  /r \cr r \lambda_- &  \lambda_+}~,
\end{equation}
where $r \equiv (1 - \epsilon)/(1+ \epsilon)$,
$\epsilon$ is the $CP$--violation parameter
\cite{PDG} and $\tau$ is the kaon proper time.
The (complex) eigenvalues of the (non--hermitian) Hamiltonian $H_{\rm free}$ are
$\lambda_{S,L} = \lambda_+ \pm \lambda_- = m_{S,L} -i\Gamma_{S,L}/2$,
where $m_{S,L}$ and $\Gamma_{S,L}$ are the masses and decay widths. The
eigenstates are 
\[
|K_{S,L} \rangle=
\frac{1}{\sqrt{2(1+|\epsilon|^2)}}\left[(1+\epsilon)|K^0\rangle\pm
(1-\epsilon)|\bar{K}^0\rangle \right] \to
\frac{1}{\sqrt{2}}\left[|K^0 \rangle\pm |\bar{K^0} \rangle\right]~, \nonumber
\]
with $|K_{S,L}\rangle\equiv |K_{S,L}(\tau=0)\rangle$.
In the final expressions, $CP$--violating effects have been ignored.
This is reasonable due to the smallness of the $CP$--violation parameter $\epsilon$
and defines the two orthogonal $CP$ eigenstates $|K_1\rangle$
($CP=+1$) and $|K_2\rangle$ ($CP=-1$). The $K_{S,L}$ time evolution
shows no oscillation between these two states and, according to
Eq.~(\ref{evol}), it is given by
\begin{equation}
\label{evo}
|K_{S,L}(\tau)\rangle= e^{-i m_{S,L} \tau}
e^{-{1 \over 2}\Gamma_{S,L} \tau} |K_{S,L}\rangle
\equiv e^{-i \lambda_{S,L} \tau} |K_{S,L}\rangle~.
\end{equation}
The $|K_{S,L}\rangle$ states define a quasi--orthonormal basis:
$\langle K_S | {K_S}\rangle = \langle K_L | {K_L}\rangle = 1$ and
$\langle K_S | {K_L}\rangle = \langle K_L | {K_S}\rangle = 2
Re\{\epsilon\}/(1 + |\epsilon |^2) \simeq 0$.
While the $\{K_S,K_L\}$ basis is useful to discuss free--space propagation, the
$CP$--basis describes weak kaon decays either into two or three final pions
from the $K_1 \simeq K_S$ or $K_2 \simeq K_L$ states, respectively.

\subsection{Two measurements: Strangeness or lifetime}
\label{measurements}

A generic neutral--kaon state is a `qubit' in a `quasispin space', i.e., a quantum
superposition of the two states of any of the previous bases,
$\{K^0,\bar K^0\}$ or $\{K_S,K_L\}$, associated to
the strangeness or lifetime quantum measurements, respectively.
The former measurement requires the introduction of a nucleonic medium in the
kaon trajectory, the latter is performed by allowing for kaon
free--space propagation. Indeed, when a kaon--nucleon reaction
occurs at a given place of the inserted medium, the distinct strong
interactions of the $S=+ 1$ and $S=- 1$ components on the bound
nucleons inside the medium project the arbitrary state of an
incoming kaon into one of the two members of the strangeness basis
\cite{AS}. The quantum number $S$ of the kaon state is determined by
identifying the products of the strangeness conserving kaon--nucleon
strong interaction.
This measurement is then analogous to the projective von Neumann measurements
with two--channel analyzers for polarized photons or Stern--Gerlach setups for
spin--$1/2$ particles. Unfortunately, the efficiency for such strangeness measurements
at moderate kaon energies as in $\phi\to K^0 \bar K^0$ and
$p \bar p\to K^0 \bar K^0$ is certainly less than what one naively expects from
the strong nature of these interactions \cite{AS,CPLEARreview}. The reason,
rather than being the difficulty in detecting the final state particles,
stems from the low probability in initiating  the strong reaction.
Indeed, the efficiency to {\it induce} a kaon--nucleon
interaction at a given time
turns out to be close to $1$ only for infinitely dense materials
or for ultrarelativistic kaons.

To measure if a kaon is propagating in free--space as a $K_S$ or $K_L$ at
a given time $\tau$, one has to allow for further propagation in free--space and then
detect at which time it subsequently decays. Kaons which show a decay vertex
between times $\tau$ and $\tau + \Delta \tau$ have to be
identified as $K_S$'s, while those decaying later than $\tau + \Delta \tau$ have to
be identified as $K_L$'s. Since there are no $K_S$--$K_L$ oscillations, such
subsequent decays do really identify the state at the desired previous time
$\tau$. The probabilities for wrong $K_S$ and $K_L$ identification are then given by
$\exp(- \Gamma_S\, \Delta \tau)$ and $1 - \exp(- \Gamma_L\, \Delta \tau)$, respectively.
Choosing  $\Delta \tau = 4.8\, \tau_S$, both misidentification
probabilities reduce to $\simeq 0.8$\%.
Since the lifetime eigenstates are not strictly orthogonal to each other,
their identification cannot be exact even in principle. However, $\epsilon$
is so small and the decay probabilities of the two
components so different ($\Gamma_S \simeq 579 \, \Gamma_L$) that the $K_S$ vs $K_L$
discrimination can effectively work \cite{eberhard,BG}.
Note also that, contrary to strangeness measurements, $K_S$ vs $K_L$ identification
is not affected by low efficiencies if detectors with
very large solid angles are used. 

\subsection{Two measurement procedures: Active or passive}

The above measurement methods are appropriate to establish Bell
inequalities and tests \cite{BBGH,BEG0,BH1,HCHSH,BHkaonicqubits}. On
the one hand, the two measurements correspond to complementary
observables, with dichotomic outcomes in both cases. On the other
hand, being performed by exerting the free will of the experimenter,
they are {\it active} measurement procedures.

However, contrary to what happens in other two--level quantum
systems, such as spin--$1/2$ particles or photons, {\it passive}
measurements are also possible for neutral kaons \cite{BGHPR} by
randomly exploiting the QM dynamics of kaon decays. To this aim, one
has to allow for complete free--space propagation and observe the
various kaon decay modes. By neglecting $CP$--violation effects,
kaon decays into two and three pions permit the identification of
$K_S$'s and $K_L$'s, respectively. Alternatively, the strangeness of
a given kaon state is measured by observing its semileptonic decays.
These decays obey the well tested $\Delta Q = \Delta S$ rule, which
allows the modes $K^0 \to \pi^- l^+ \nu_l $ and $\bar{K^0} \to \pi^+
l^- \bar \nu_l $, with $l = e, \mu $, but forbids decays into the
respective charge--conjugate final states \cite{PDG}. These
procedures for the passive $K_S$ vs $K_L$ and $K^0$ vs $\bar{K^0}$
discriminations are unambiguous in the approximations given by
$CP$--conservation and the $\Delta Q = \Delta S$ rule, respectively.
However, the experimenter has no control on the time when the
measurement occurs, nor on the basis in which it is performed, in
contrast with the previous active procedure. As a result, the so
called Kasday construction \cite{Kasday} invalidates Bell inequality
tests performed with such passive measurements \cite{BEG0}.

\subsection{Two--kaon systems: Entanglement}
The simplest and most often discussed bipartite state is the spin
singlet state consisting of  two photons or two spin--1/2 particles, as first
proposed by D.~Bohm \cite{Bohm}. Let us then first
consider the analogous two--kaon entangled state 
 \cite{BHkaonicqubits,bn,gigo,Bert}. 
Both $\phi$--resonance decays \cite{daphne} and  $s$--wave
proton--antiproton annihilations \cite{CPLEARreview} produce 
the $J^{PC}=1^{--}$ initial state:
\begin{eqnarray}
\label{entangled}
|\phi(\tau =0)\rangle  &=&  \frac{1}{\sqrt 2}\left\{
|K^0\rangle_l |\bar{K}^0\rangle_r - |\bar{K}^0\rangle_l
|K^0\rangle_r\right\} \\
 &=&  \frac{1}{\sqrt 2}\frac{1+|\epsilon|^2}{|1-\epsilon^2|}\left\{
|K_L\rangle_l |K_S\rangle_r - |K_S\rangle_l |K_L\rangle_r\right\}~, \nonumber
\end{eqnarray}
$l$ and $r$ denoting the `left' and `right' directions of
motion of the two separating kaons. The weak, $CP$--violating effects enter
only in the last equality.

After production, the left and right moving kaons evolve
according to Eq.~(\ref{evo}) up to  left-- and right--times $\tau_l$ and
$\tau_r$, respectively. Once normalizing to surviving kaon pairs,
this leads to the $\Delta \tau=\tau_l-\tau_r$ dependent state
\begin{equation}
\label{timeentangled}
|\phi(\Delta\tau)\rangle = \frac{1}{\sqrt {1+e^{\Delta\Gamma
\Delta t}}}\biggl\lbrace |K_L\rangle_l|K_S\rangle_r - e^{i \Delta m \Delta\tau}
e^{{1 \over 2} \Delta
\Gamma \Delta\tau}|K_S\rangle_l|K_L\rangle_r\biggr\rbrace~,
\end{equation}
in the lifetime basis, with $\Delta m \equiv m_L -m_S$ and
$\Delta \Gamma \equiv \Gamma_L - \Gamma_S$.
For $\Delta \tau = 0$ this state shows maximal entanglement in both
lifetime and strangeness.

This state (\ref{timeentangled}) is analogous to the polarization--entangled
two--photon [idler ($i$) plus signal ($s$)] state used in different optical tests:
\begin{equation}
|\Psi \rangle = \frac{1}{\sqrt{2}}
\left\{ |V\rangle_i |H\rangle_s - e^{i \Delta \phi}|H\rangle_i |V\rangle_s \right\} ,
\end{equation}
where $\Delta \phi$ is an adjustable relative phase and $V$ and $H$
refer to vertical and horizontal polarizations. For entangled kaons, 
$\Delta m$ plays the role of $\Delta \phi$ and induces strangeness
oscillations in time which can be used to mimic the different
orientations of polarization analyzers in photonic Bell--tests
\cite{bn,gigo}. Note, however, that the two terms in the photonic
state have the same weight but that this is not the case in the
two--kaon states (\ref{timeentangled}). 

The entanglement of the kaonic state (\ref{timeentangled}) has been tested
experimentally at CPLEAR over macroscopic distances 
and using active strangeness measurements \cite{CPLEARreview}. The
non--separability of this state has also been observed at the Daphne
$\Phi$--factory using passive measurements \cite{preDiDo}; with some
modification of the set--up, this could be possible by active
measurements too.

In Ref.~\cite{zetaberthies} the authors analyzed the possibility of
a spontaneous wave--function factorization, which was proposed by
Schr\"odinger and Furry in $1935$. If a factorization in the $\{K_S,
K_L\}$ basis is assumed, the QM interference term is
simply multiplied by $(1-\zeta)$
\[
P_\zeta \left[K^0(\tau_l), K^0(\tau_r)\right] = {1 \over 4} 
\left\{1- 2(1- \zeta) \frac{\cos(\Delta m \Delta \tau ) 
e^{-(\Gamma_S+\Gamma_L)(\tau_l + \tau_r)/2}}
{e^{-\Gamma_S \tau_l -\Gamma_L \tau_r} + e^{-\Gamma_L \tau_l+\tau_r)}}\right\}~, \nonumber
\]
where $\zeta$ is the `decoherence parameter', $0 \le \zeta \le
\zeta_{\textrm{Schr\"odinger--Furry}} = 1$, characterizing the
strength of the interaction of the entangled state with the
environment. If a factorization in the $\{K^0, \bar K^0\}$ basis is
assumed then it is a bit more complicated. Now one can compare these
models with the experimental data from the CPLEAR experiment
\cite{CPLEARreview} and a recent experiment of the
KLOE--collaboration at Daphne \cite{preDiDo}
\begin{eqnarray}
\zeta_{K_S,K_L} &=& 0.13 \pm 0.16 \;  \cite{CPLEARreview};  \; \;
0.018 \pm 0.040_{\rm stat} \pm 0.007_{\rm syst} \;\cite{preDiDo} \nonumber \\
\zeta_{K^0,\bar{K^0}} &=& 0.4 \pm 0.7\; \cite{CPLEARreview} ; \; \;
(0.10 \pm 0.21_{\rm stat} \pm 0.04_{\rm syst}) \cdot 10^{-5}
\;\cite{preDiDo}.
\end{eqnarray}
The results in the $\{K^0\bar K^0\}$ basis of the KLOE experiment
benefits from large cancellations between the interference term and
the two terms that occur for the $CP$ suppressed final state
$\pi^+\pi^-$. 

\section{Quantitative Complementarity}
\label{quco}

As it is well known since long time,
the observation of an interference pattern and the
acquisition of `which way' information are mutually excluded 
in interferometric devices. However,
quantitative statements of this complementarity principle have become
available only recently.

The quality of an interference pattern can be quantified in
terms of the `fringe visibility'
${\cal V}_0=(I_{\rm max} - I_{\rm min})/(I_{\rm max} + I_{\rm min})$,
where $I_{\rm max, min}$ stand for the maximum and minimum measured intensities.
In general, for a two--path interferometer the intensity is
$I(\phi ) \propto (1 + {\cal V}_0 \cos \phi )$,
where $\phi$ is the phase difference between the two paths.
The amount of which--path information is given by the `predictability' \cite{GY}
${\cal P}  \equiv |w_I - w_{II}|$,
where $w_{I (II)}$ is the probability to take the
interferometric path $I$ ($II$).
Then, complementarity can be expressed in the quantitative form
\begin{eqnarray}
\label{QC}
{\cal P}^2 + {\cal V}_0^2  \le 1~,
\end{eqnarray}
where the equal sign is valid for pure states.

The evolution of a single kaon state can be interpreted
in terms of this duality relation \cite{BGH04,BGH2}.
By normalizing to kaons surviving up to time $\tau$ and neglecting
$CP$--violation effects, the time evolution of an initial
$K^0$ state (an analogous discussion holds for initial $\bar{K^0}$'s)
can be written as
\[
|K^0 (\tau )\rangle = {1 \over \sqrt{2}}
\left[ {1 + e^{-{1\over 2}\Delta \Gamma \tau} e^{-i \Delta m \tau}
\over \sqrt{1 + e^{-\Delta \Gamma t}}} |K^0\rangle +
       {1 - e^{-{1\over 2}\Delta \Gamma \tau} e^{-i \Delta m\tau}
\over \sqrt{1 + e^{-\Delta \Gamma t}}} |\bar{K^0}\rangle \right]~. \nonumber
\]
The strangeness oscillations in $\tau$ of $|{K^0} (\tau)\rangle$
are characterized by the phase
$\phi(\tau)=\Delta m\, \tau$ and by a time dependent visibility and path
predictability
\begin{equation}
\label{VK}
{\cal V}_{0} (\tau)=\cosh^{-1} \left(\Delta \Gamma\, \tau/2\right)~,\,\,\,
{\cal P} (\tau) = \tanh \left(\Delta \Gamma\, \tau/2 \right)~,
\end{equation}
which satisfy Eq.~(\ref{QC}), ${\cal P}^2(\tau) + {\cal V}^2_0 (\tau) =1$.
This clearly shows the interferometric behaviour of
neutral kaon evolution, where the $K_S$ and $K_L$ 
components play the role of the two interferometric paths \cite{BGH04,BGH2}.

\section{Quantum Eraser}
\label{qe}
The quantum eraser is a subtle phenomenon originally proposed by Scully and
Dr$\ddot{\rm u}$hl \cite{scully82} and recently reviewed by Aharonov and Zubairy
\cite{AZ}. It has been demonstrated  in several atomic and photonic experiments
but it can be performed, with some advantages, by using neutral kaons
\cite{BGH1}.

In this type of analyses one considers variations of the basic double--slit experiment.
In a two--way experiment, interference patterns are observed if and only if
it is impossible to know, \emph{even in principle}, which way the
particle took. Interference disappears if there is a way to know
---through a \emph{quantum marking} procedure--- which
way the \emph{object} particle took;
whether or not the outcome of the corresponding `which way'
observation, performed on the \emph{meter} particle,
is actually read out, it does not matter: interference is in any way
lost. 
For a two--particle entangled state, if
the path of one member is marked,
information on the path taken by its entangled partner is in principle available
and no interference fringes can be observed. But, if that `which way' mark
is erased by means of a suitable measurement
---\emph{quantum erasure}---,
interferences reappear in joint detection events.

For neutral kaons, the phenomenon of $K^0$--$\bar{K^0}$ oscillations
plays the role of the standard interference fringes.
Similarly, the $K_S$ and $K_L$ states,
showing a distinct propagation in free--space, are the analogs
of the two separated trajectories in interferometers.

Consider the two--kaon entangled state (\ref{timeentangled}).
The \emph{object} kaon flying to the left hand side is always measured
\textit{actively} in the strangeness basis.
This measurement is performed by placing the strangeness detector at
different points of the left trajectory, thus scanning for
oscillations along a certain $\tau_l$ range. The kaon flying to the
right hand side, the \emph{meter}, is always measured
\textit{actively} at a fixed time $\tau_r^0$, either in the
strangeness or in the lifetime basis. In the latter case we obtain
full `\textit{which width}' information for the object kaon
---analogously to the `\textit{which way}' information in a double
slit; consequently, no interference in the meter--object joint
detections can be observed. This can be  seen from
Eq.~(\ref{timeentangled}) leading to the
non--oscillating joint probabilities:
\begin{eqnarray}
\label{probS}
&&P\left[K^0(\tau_l),K_S(\tau^0_r)\right]=P\left[\bar{K}^0(\tau_l),K_S(\tau^0_r)\right]
=\frac{1}{2\left(1+e^{\Delta\Gamma\Delta\tau}\right)} , \\
\label{probL}
&&P\left[K^0(\tau_l),K_L(\tau^0_r)\right]=P\left[\bar{K}^0(\tau_l),K_L(\tau^0_r)\right]
=\frac{1}{2\left(1+e^{-\Delta\Gamma\Delta\tau}\right)} .
\end{eqnarray}

However, the possibility to obtain `which width' information can be
precluded by quantum erasure, i.e., by measuring strangeness on
the meter kaon thus making its $K_S$ or $K_L$ `mark' inoperative.
The joint probabilities:
\begin{eqnarray}
\label{lSprob}
P\left[K^0(\tau_l),K^0(\tau^0_r)\right]=P\left[\bar{K}^0(\tau_l),\bar{K}^0(\tau^0_r)\right]
= \frac{1}{4}\left[1-{\cal V}(\Delta\tau) \cos(\Delta m\, \Delta \tau)\right]\,\,\, && \\
\label{uSprob}
P\left[K^0(\tau_l),\bar{K}^0(\tau^0_r)\right]=P\left[\bar{K}^0(\tau_l),K^0(\tau^0_r)\right]
= \frac{1}{4}\left[1+{\cal V}(\Delta\tau) \cos(\Delta m\, \Delta
\tau)\right]\,\,\, &&
\end{eqnarray}
then show $\tau_l$--dependent strangeness oscillations with visibility
${\cal V}(\Delta\tau)=\cosh^{-1}(\Delta\Gamma\Delta\tau/2)$.

The previous kaon quantum eraser has some
advantages as compared to the standard photonic case.
With kaon pairs, the two `paths' ($K_S$ and
$K_L$ propagation) are already present and automatically `marked' ($\Gamma_S
>> \Gamma_L$) from the very beginning simplifying the state preparation.  
From a more theoretical point of view, one notes that
the  oscillating probabilities
(\ref{lSprob}) and (\ref{uSprob}) are even functions of
$\Delta \tau$. Which measurement, left or right, is first performed is then
irrelevant and the quantum eraser can be
operated in the so--called `delayed choice' mode \cite{BGHPR}.
In this mode, the decision to observe or not strangeness oscillations when scanning
the object kaon can be taken once this kaon has already been detected by
performing a future measurement of strangeness or lifetime on the
meter kaon.

These latter comments also add some light to the very nature of the quantum eraser
working principle: the way in which joint detection events are {\it classified} according
to the available information. In the `delayed choice' mode, a series of strangeness
measurements is performed at different $\tau_l$ times on the object kaons and the
corresponding outcomes are recorded. Later one can measure  either lifetime or strangeness
on the corresponding meter partner and, only then, full information allowing for a
definite sorting of each pair is available. Choosing to perform strangeness
measurements on the meter kaons amounts to completely erase the `which width' information
on each pair in such a way that  oscillations and
complementary anti--oscillations appear in the corresponding subsets. The alternative
choice of lifetime measurements on meter kaons, instead, does not offer the
possibility to classify the events in oscillatory subsets as before.

\section{Bell Inequalities Tests}
\label{bit} In this Section we briefly discuss some promising ideas
to test Bell inequalities with kaons, i.e., testing Local Realism (LR)
vs QM. Further discussions on this subject  can be found in 
Refs.~\cite{BH1,HCHSH,Bert,BEG,berthies} and references
therein.

The derivation of the CHSH--Bell  inequality for neutral kaons
\cite{BH1} follows 
the original proof by Clauser {\it et al.} 
in 1969, Ref.~\cite{CHSH}, which is an
extension of Bell's proof  under more realistic
assumptions. One finds 
\begin{eqnarray}\label{chsh}
S_{k_n,k_m,k_{n'},k_{m'}}(t_1,t_2,t_3,t_4)&=&\left|
E_{k_n,k_m}(t_1,t_2)-E_{k_n,k_{m'}}(t_1,t_3)\right|\nonumber\\
&&\hspace{-1cm} +|E_{k_{n'},k_{m}}(t_4,t_2)+E_{k_{n'},k_{m'}}(t_4,t_3)|\leq 2~.
\end{eqnarray}
For kaons, each experimenter has choices of two 
different kinds: one choice refers to the `quasispin' state 
and the other to the time the kaon
propagates until the measurement. 
The situation is then somehow more interesting than in the photonic case 
but also more involved because  of the kaon evolution and decay. 
As in the usual photon setup,
Alice and Bob can choose among two settings, i.e., Alice: $\{(k_n,
t_1); (k_{n'}, t_4)\}$ and Bob: $(\{(k_m, t_2); (k_{m'}, t_3)\})$.
The expectation value $E_{k_n,k_m}(t_1,t_2)$ denotes then that Alice
chooses to measure the quasispin $k_n$ at time $t_1$ on the kaon
propagating to her side and Bob chooses to measure $k_m$ at time
$t_2$ on his kaon. 

If we fix the quasispins to  the strangeness eigenstates, we
obtain a Bell inequality depending on four times, which is in close
analogy to photonic cases. But, surprisingly,  working with the maximally entangled state
(\ref{entangled}) no violation of the inequality (\ref{chsh}) can be obtained 
\cite{BBGH,BH1}. By contrast, for  non--maximal entangled states 
violations of this Bell inequality (up to $S=2.159$) can be found,
see Ref.~\cite{HCHSH}. Discussions with the experimenters at
Daphne on the feasibility of these states are ongoing.

Apart from the previous states of kaons, other non--maximally
entangled states are of interest for Bell tests. 
In Ref.~\cite{BG} we have proposed the state  
\begin{equation}
\label{stateN}
|\Phi\rangle =
{1 \over \sqrt{2 + |R|^2}}
\left[|K_S\rangle |K_L\rangle  - |K_L\rangle |K_S\rangle
+ R |K_L\rangle |K_L\rangle \right]~,
\end{equation}
which can be produced at a $\Phi$--factory with the use of a kaon
regenerator. Here, $R \equiv  -\eta
\exp\{[-i\Delta m +{1 \over 2}(\Gamma_S - \Gamma_L)]T\}$, $\eta$
being the regenerator parameter.
The non--maximally entangled state $\Phi$ describes all kaon
pairs with both left and right partners surviving up to a common proper time $T$,
with $\tau_S << T << \tau_L \simeq 579\, \tau_S$.

This state can be conveniently used
for Bell--type tests. Following the approach of
Ref.~\cite{BG}, for each kaon on each beam at time $T$ we consider either a
strangeness or a lifetime measurement.
With the strategy of subsection \ref{measurements} for lifetime
measurements, requiring an extra interval time $\Delta T=4.8\, \tau_S$ after $T$,
care has to be taken to choose $T$ large enough to guarantee the space--like
separation between left and right measurements.
For kaon pairs from $\phi$ decays,
this implies $T > 1.77\, \Delta T$.

The following Clauser--Horne (CH) inequalities have been derived
under the assumption of perfectly efficient experimental apparata
(fair sampling hypothesis) in Ref.\cite{BG}:
\begin{eqnarray}
\label{CHsecondo}
&&\frac{P(\bar{K^0} ,K_L)-P(\bar{K^0} ,\bar{K^0} ) + P(K_S, \bar{K^0} )
+P(K_S ,K_L)}{P(K_S ,*) + P(*,K_L)} \leq 1 ,\\
&&\frac{P(\bar{K^0},K_S) -P(\bar{K^0} ,\bar{K^0} ) + P(K_L,\bar{K^0})
+P(K_L,K_S)}{P(*,K_S) - P(K_L,*)} \leq 1, \nonumber
\end{eqnarray}
where, for instance, $P(K_S,*) \equiv P(K_S,K^0) + P(K_S,\bar{K}^0)$.
Note that each one of the two
inequalities follows from the other by just inverting left and right
measurements on the left--right asymmetric state (\ref{stateN}).

By substituting the QM predictions in the
inequalities (\ref{CHsecondo}), one finds:
\begin{equation}
\label{QM}
{2 - {\cal{R}}e\, R +{1\over 4}|R|^2 \over 2 +|R|^2} \leq 1, \; \;
{2 + {\cal{R}}e\, R +{1\over 4}|R|^2 \over 2 +|R|^2} \leq 1,
\end{equation}
whose only difference is the sign affecting the linear term in ${\cal{R}}e\, R$.
According to this sign, one of these two inequalities is
violated if $|{\cal{R}}e\, R| \geq 3|R|^2/4$. The greatest violation occurs for a
purely real value of $R$, $|R| \simeq 0.56$, for which one of the two
ratios in Eq.~(\ref{QM}) reaches the value 1.14.
This 14~\% violating effect predicted by QM opens up the possibility 
of refuting LR models modulo the fair sampling hypothesis.

We conclude our review by discussing a proposal that does not assume
auxiliary hypotheses 
going beyond the reality and locality requirements. 
In our opinion, it represent an interesting
attempt for a loophole--free test of LR vs QM with
neutral kaons.
It is based on Hardy's proof of Bell's theorem \cite{Ha93} without inequalities 
and it has been applied in Ref.~\cite{BG1} to the non--maximally
entangled state (\ref{stateN}). This considerably
improves the analysis of Ref.~\cite{BG}.
Indeed, Hardy's non--locality proof can be translated into a Bell inequality
\cite{BEG} which could discriminate between LR and QM
if the detection efficiencies for strangeness and lifetime
measurements at disposal are high enough.

Let us first concentrate on the `non--locality without inequalities'
proof of Ref.~\cite{BG1}. Neglecting $CP$--violation and
$K_L$--$K_S$ misidentifications, from state (\ref{stateN}) with
$R=-1$ (Hardy's state) one obtains the QM predictions:
\begin{eqnarray}
\label{non-zero}
P_{\rm QM}(K^0,\bar{K}^0)&=&\eta\, \bar \eta/12 , \\
\label{zero1}
P_{\rm QM}(K^0,K_L)&=&0 , \\
\label{zero2}
P_{\rm QM}(K_L,\bar{K}^0)&=&0 , \\
\label{quasi-zero}
P_{\rm QM}(K_S,K_S)&=& 0 ,
\end{eqnarray}
where $\eta$ ($\bar \eta$) is the overall efficiency
for $K^0$ ($\bar K^0$) detection.
It is found that the necessity to reproduce, under LR, equalities
(\ref{non-zero})--(\ref{zero2}) requires
$P_{\rm LR}(K_S,K_S) \geq P_{\rm LR}(K^0,\bar{K}^0)=\eta\, \bar \eta/12>0$,
which contradicts Eq.~(\ref{quasi-zero}).
In principle, this allows for a test of LR vs QM without inequalities.
However, since $K_L$--$K_S$ misidentifications
(due to the finite value of $\Gamma_S/\Gamma_L\simeq 579$) preclude an ideal
lifetime measurement even when the detection efficiency $\eta_\tau$ for the kaon
decay products is $100$\%, the above proposal must be reanalised
paying attention to these inefficiencies \cite{BEG}.

Retaining the $K_S$--$K_L$ misidentifications effects,
the results (\ref{non-zero})--(\ref{quasi-zero}) are replaced by
(see the Appendix of Ref.~\cite{BEG} for details):
\begin{eqnarray}
\label{SS0}
P_{\rm QM}(K^0,\bar{K}^0) &=& \eta \bar \eta/12 , \\
\label{SL0}
P_{\rm QM}(K^0,K_L) &=&
6.77 \times 10^{-4}\, \eta \, \eta_{\tau} , \\
\label{LS0}
P_{\rm QM}(K_L,\bar K^0) &=&
6.77 \times 10^{-4}\, \bar \eta \, \eta_{\tau} , \\
\label{LL0}
P_{\rm QM}(K_S,K_S)
&=& 1.19 \times 10^{-5}\, \eta_\tau^2.
\end{eqnarray}
In the standard Hardy's proof  \cite{Ha93},
the probabilities corresponding to our (\ref{SL0})--(\ref{LL0})
are perfectly vanishing. In our case they are very small but not zero.
Nevertheless, this does not prevent from deriving a contradiction between LR
and QM. 
One has to use the  Eberhard inequality \cite{BEG,Eb}:
\begin{equation}
\label{hardy}
H\equiv \frac{P(K^0,\bar{K}^0)}{P(K^0,K_L)+ P(K_S,K_S) + P(K_L,\bar{K}^0)+
P(K^0,F)+P(F,\bar K^0)} \leq 1~,
\end{equation}
where the argument $F$ refers to failures in lifetime detection, and,
in QM:
\begin{eqnarray}
\label{extra1}
P_{\rm QM}(K^0,F)= \frac{1}{6} \eta \left(1-\eta_{\tau} \right),  \;\;
P_{\rm QM}(F,\bar K^0)=\frac{1}{6} \bar \eta \left(1-\eta_{\tau}
\right) .
\end{eqnarray}
Note that the use of an inequality also allows for
deviations, existing in real experiments, in the value of
$R$ required to prepare Hardy's state.
It is important to stress that the inequality (\ref{hardy}) has been
obtained {\it without invoking supplementary assumptions} on undetected events.
From this inequality one obtains the restrictions on the efficiencies
$\eta$, $\bar \eta$ and $\eta_{\tau}$ required for a loophole--free
experiment.

To discuss the feasibility of such an experiment, let us start considering
a few ideal cases. Assume first that perfect discrimination
between $K_S$ and $K_L$ were possible;
one could then make a conclusive test of LR for any nonvanishing
values of $\eta$ and $\bar \eta$:
$H_{\rm QM}\to \infty$,
$\forall\, \eta, \bar \eta\neq 0$. In a second ideal case with no undetected
events, i.e., $\eta=\bar \eta=\eta_{\tau}=1$, the inequality is strongly
violated by QM, $H_{\rm QM}\simeq 60.0$,
even if one allows for unavoidable $K_S$ and $K_L$ misidentifications. Finally,
assuming that only the detection efficiency of kaon decay products is ideal
($\eta_{\tau}=1$), for $\eta=\bar \eta$ ($\eta=\bar \eta/2$) Eberhard
inequality is contradicted by QM for $\eta> 0.023$ ($\eta> 0.017$).

More realistic situations, with small and possibly achievable values of $\eta$ and
$\bar \eta$, must be considered. According to the results of Ref.~\cite{BEG},
to have a loophole--free test,
this implies that we have to consider large decay--product
detection efficiencies such as $\eta_{\tau}\simeq 0.98$, for which
$\eta$ and $\bar \eta$ can be lowered to about 0.06
(see Fig.~1 of Ref.~\cite{BEG}).
The values of $\eta_\tau$, $\eta$ and $\bar \eta$ required by the test
proposed in Ref.~\cite{BEG} seem to be not far from the present experimental capabilities.



\end{document}